\documentclass[fleqn,traditabstract]{aa}
\usepackage{graphicx}
\usepackage{amsmath}
\usepackage[varg]{txfonts}
\usepackage[sort]{natbib}
\usepackage{color}
\bibpunct{(}{)}{;}{a}{}{,}

\newcommand{\kms}{\ensuremath{\text{km}\,\text{s}^{-1}}}
\newcommand{\Msun}{\ensuremath{\text{M}_{\odot}}}
\newcommand{\Lsun}{\ensuremath{\text{L}_{\odot}}}

\newcommand{\hb}{\ensuremath{\text{H}\beta}}
\newcommand{\ha}{H$\alpha$}
\newcommand{\teff}{\ensuremath{T_\text{eff}}}
\newcommand{\Te}{\ensuremath{T_\text{e}}}
\newcommand{\Ne}{\ensuremath{n_\text{e}}}
\newcommand{\ccm}{\ensuremath{\text{cm}^{-3}}}
\newcommand{\SB}{\ensuremath{\text{erg}\,\text{cm}^{-2}\,\text{s}^{-1}\,\text{arcsec}^{-2}}}
\newcommand{\Fl} {\ensuremath{\text{erg}\,\text{cm}^{-2}\,\text{s}^{-1}}}
\newcommand{\FlA}{\ensuremath{\text{erg}\,\text{cm}^{-2}\,\text{s}^{-1}\,\text{\AA}^{-1}}}
\newcommand{\xx}{\ensuremath{\phantom{0}}}

\def\rPT{PT05}
\def\rS{S05}
\def\rR{R02}
\def\rJ{J02}
\def\rJs{J06}
\def\rT{T01}
\def\rTo{T04}
\def\BoBn{BoBn-1}
\def\rP{Paper~VII}
\def\rPtwo{Paper~II}
\def\obj{PN\,G135.9+55.9}
\def\ugd{\ensuremath{_{\text{GD}}}}

\begin{document}

\title{The evolution of planetary nebulae}
\subtitle{VI.\ On the chemical composition of the metal-poor {\obj}\thanks{Based in part on observations collected at the Centro Astron\'omico Hispano Alem\'an (CAHA), operated jointly by the Max-Planck-Institut f\"ur Astronomie and the Instituto de Astrofisica de Andalucia (CSIC).}}

\titlerunning{The evolution of planetary nebulae VI.}

\author{C.\ Sandin \and R.\ Jacob \and D.\ Sch\"onberner \and M.\ Steffen \and M.\ M.\ Roth}

\institute{Astrophysikalisches Institut Potsdam, An der Sternwarte 16, D-14482 Potsdam, Germany\\\email{csandin@aip.de, rjacob@aip.de, deschoenberner@aip.de, msteffen@aip.de, mmroth@aip.de}}

\offprints{csandin@aip.de}

\date{Received February 5, 2009 / Accepted December 22, 2009}

\abstract{The actual value of the oxygen abundance of the metal-poor planetary nebula \object{\obj} has frequently been debated in the literature. We wanted to clarify the situation by making an improved abundance determination based on a study that includes both new accurate observations and new models. We made observations using the method of integral field spectroscopy with the PMAS instrument, and also used ultraviolet observations that were measured with HST-STIS. In our interpretation of the reduced and calibrated spectrum we used for the first time, recent radiation hydrodynamic models, which were calculated with several setups of scaled values of mean Galactic disk planetary nebula metallicities. For evolved planetary nebulae, such as \object{\obj}, it turns out that departures from thermal equilibrium can be significant, leading to much lower electron temperatures, hence weaker emission in collisionally excited lines. Based on our time-dependent hydrodynamic models and the observed emission line $[\ion{O}{iii}]\,\lambda5007$, we found a very low oxygen content of about 1/80 of the mean Galactic disk value. This result is consistent with emission line measurements in the ultraviolet wavelength range. The C/O and Ne/O ratios are unusually high and similar to those of another halo object, \BoBn.}


\maketitle

\section{Introduction}\label{sec:introduction}
The stellar-like object \object{SBS 1150+599A} from the Second Byurakan Survey \citep{Ba:97} has been spectroscopically identified by \citet[hereafter {\rT}; \citealt{GaSt:99}]{ToStCh.:01} to be an old planetary nebula (PN) of the Galactic halo and was thereafter renamed \object{\obj}. The same authors perform an abundance study based on photoionization models and come to the conclusion that this particular object has the lowest oxygen abundance known so far for planetary nebulae, viz.\ below about 1/100 of the solar value.

This object appears to be a challenge for any detailed spectroscopic analysis since very few emission lines are detectable in the optical wavelength range. Because of a lack of suitable lines, it is impossible to make a direct plasma diagnostic, and it is also difficult to constrain photoionization models such that meaningful abundances will emerge. The analysis is, moreover, hampered since this object is faint, with a mean {\hb} surface brightness of only $\sim\!5\!\times\!10^{-16}$\,{\SB} (adopting a diameter of 5{\arcsec} and a total {\hb} flux of  $1.9\!\times\!10^{-14}$\,{\Fl}; see Table \ref{sandint1}), making it difficult to accurately measure lines close to the 1\% level of {\hb} (corresponding to $\simeq\!2\!\times\!10^{-16}$\,{\Fl}).

These observational difficulties have led to a dispute about whether \object{\obj} is as extremely metal-deficit as {\rT} claim. A critical point in this discussion is the determination of the stellar temperature from the ionization balance, using the line ratio of $[\ion{Ne}{v}]\,\lambda3426$ and $[\ion{Ne}{iii}]\,\lambda3869$ in the nebula. The line strength of $[\ion{Ne}{iii}]\,\lambda3869$ is uncertain, but decisive when fixing the effective temperature ({\teff}) of the ionizing source. \citet[][hereafter \rR]{RiToSt.:02} and \citet[][hereafter \rJ]{JaFeCl.:02} find a rather low $\teff\!\simeq\!100\,000$\,K, hence low oxygen abundances of less than 1/100 solar from the weak $[\ion{O}{iii}]\,\lambda5007$ line, the only oxygen line that is observed in the optical.

\citet[hereafter \rTo]{ToNaRi.:04} conclude, by means of optical and ultraviolet (UV) spectra (FUSE), that the central ionizing source of \object{\obj} must be a very hot ($\teff\!\approx\!120\,000$\,K) pre-white dwarf of rather low mass ($\approx\!0.55$\,\Msun), which resides in a short-period binary system with a more massive companion \citep[$P\!=\!3.92$\,h,][]{NaToRi.:05}. Recently \citealt{ToToNa.:07}, based on X-ray observations, have estimated that the temperature of this massive component is very high, $\teff\!\simeq\!170\,000\,$K, while the less massive optical-UV component is cooler, $\teff\!\simeq\!58\,000\,$K.

\citet[][hereafter \rPT]{PeTs:05}, moreover, critically examine the situation and come to the conclusion that all evidence favors a higher temperature of the central source, viz.\@ $\teff\!\simeq\!130\,000$\,K, and that the claimed strength of $[\ion{Ne}{iii}]\,\lambda3869$ most likely is wrong.  The oxygen abundance would then be 1/30--1/15 solar and not as extreme as previously estimated. \citet[hereafter \rJs]{JaGaBo.:06} also find a higher stellar temperature, using the much stronger UV emission lines of highly ionized nitrogen and carbon to constrain \teff. \citet[hereafter \rS]{StToRi.:05} include UV lines in their study (likely using the same HST-STIS data as \rJs) and conclude that the oxygen abundance is lower than what {\rPT} find, 1/130--1/40.

Our aim was to better measure the nebular emission line spectrum, which is why we performed new observations using the integral-field spectrograph PMAS. With such a spectrum we could determine a reliable line strength, or an upper limit, of $[\ion{Ne}{iii}]\,\lambda3869$. We begin in Sect.~\ref{sec:obsdata} with a description of our observations and data processing, we then present our results in Sect.~\ref{sec:results}. We describe the physical setup of our radiation hydrodynamic models, how we estimate abundances and compare observations with our models, in Sect.~\ref{sec:dataanalysis}. In Sect.~\ref{sec:discussion} we discuss non-equilibrium effects and compare our abundances with values of previous studies found in the literature. We close the paper with our conclusions in Sect.~\ref{sec:conclusions}.

\section{Observations and data reduction}\label{sec:obsdata}
Our observations were made with the 3.5\,m telescope at Calar Alto using the lens array (LARR) integral field unit (IFU) of the PMAS instrument \citep{RoKeFe.:05}. The V600 grating was used to cover the spectral interval 3490--5150\,{\AA}; at a dispersion of 0.81\,\AA\,pixel$^{-1}$ and a resolving power of ${R=1340}$. The wavelength range was chosen to cover the Balmer lines {\hb} and H$\gamma$, as well as [\ion{O}{iii}]$\,\lambda\lambda\,4959,\,5007$, \ion{He}{ii}$\,\lambda4686$, and [\ion{Ne}{iii}]$\,\lambda3869$. The LARR IFU, furthermore, holds $16\!\times\!16$ separate fibers, where each fiber represents a spatial element on the sky. We used the 0\farcs5 sampling mode where every pointing with the IFU covers an area of ${8\arcsec\times8\arcsec=64\arcsec^2}$ on the sky. Any number of spatial elements can be co-added to create a final spectrum.

Two 2700\,s exposures, which were both centered on {\obj}, were taken at an airmass of 1.09--1.12 on 2007 February 13. Two 2700\,s exposures were also taken at an airmass of 1.12--1.19 on 2007 February 14. These two latter exposures were offset by 5{\arcsec} E from the first two exposures. One additional 1800\,s exposure was taken at an airmass of 1.09 and an offset of 5{\arcsec} W in the second night. Weather conditions were less than optimal, and the seeing was 1\farcs4--1\farcs7. In addition to the science exposures continuum and arc lamp flatfields were taken, as well as spectrophotometric standard-star exposures of G191-B2B. In order to correct for a varying fiber-to-fiber transmission sky flats were taken at the beginning of the second night.

Reducing the data we used the tool \textsc{P3d\_online} that is a part of the PMAS \textsc{P3d} pipeline \citep{TBe:02,RoKeFe.:05}. At first the bias level was subtracted and cosmic-ray hits removed \citep[cf.][]{SaScRo.:08}. Second, a trace mask was generated from an internal continuum calibration lamp exposure, identifying the location of each spectrum on the CCD along the direction of cross-dispersion. Third, a dispersion mask for wavelength calibration was created using an arc-exposure. In order to minimize effects due to a significant flexure in the instrument all continuum and arc lamp exposures were taken within two hours of the respective science exposure. Fourth, a correction to fiber-to-fiber sensitivity variations was applied by dividing with an extracted and normalized sky flat-field exposure. In this process the data was changed from a CCD-based format to a row-stacked-spectra format. As a final step flux calibration was performed in \textsc{iraf} using standard-star exposures. We did not correct for differential atmospheric refraction since we, due to poor spatial resolution and high seeing, have co-added the flux of 147 spatial elements. These elements fully cover the object with an area of $36.75\arcsec^2$. In order to avoid the periodic line shift of stellar absorption lines of the central star (CS) we made a second spectrum where we did not add the nine spatial elements, which were the closest to the CS at $\lambda_{\text{H}\beta}$ (we also used fewer elements on the outer boundary); this latter spectrum covers an area of $30.75\arcsec^2$.

{\hb} is blended by the helium Pickering line $\ion{He}{ii}\,\lambda4859$, and the resulting line intensity is hereby about 5\% too high (cf.\@ {\rJ}; \rPT). The other Balmer lines are likewise blended by other helium Pickering lines. Since we did not include the CS in our final spectrum we have not corrected for underlying stellar absorption as {\rPT} do. The discussion of the dereddening performed by {\rR} remains inconclusive, with the result of a somewhat uncertain and very low extinction towards \object{\obj}.  Also {\rTo} find a low interstellar extinction towards this object ($E(B\!-\!V)\!=\!0.04$) that is mainly based on the $N_{\text{H\,I}}$ column density they derive from FUSE spectra. Owing to the large differences of the line strengths found by different observers (see below), however, any (rather small) correction due to dereddening appears unimportant at this point of our discussion.

\section{Results for {\obj}}\label{sec:results}
\begin{figure*}
\centering
\includegraphics{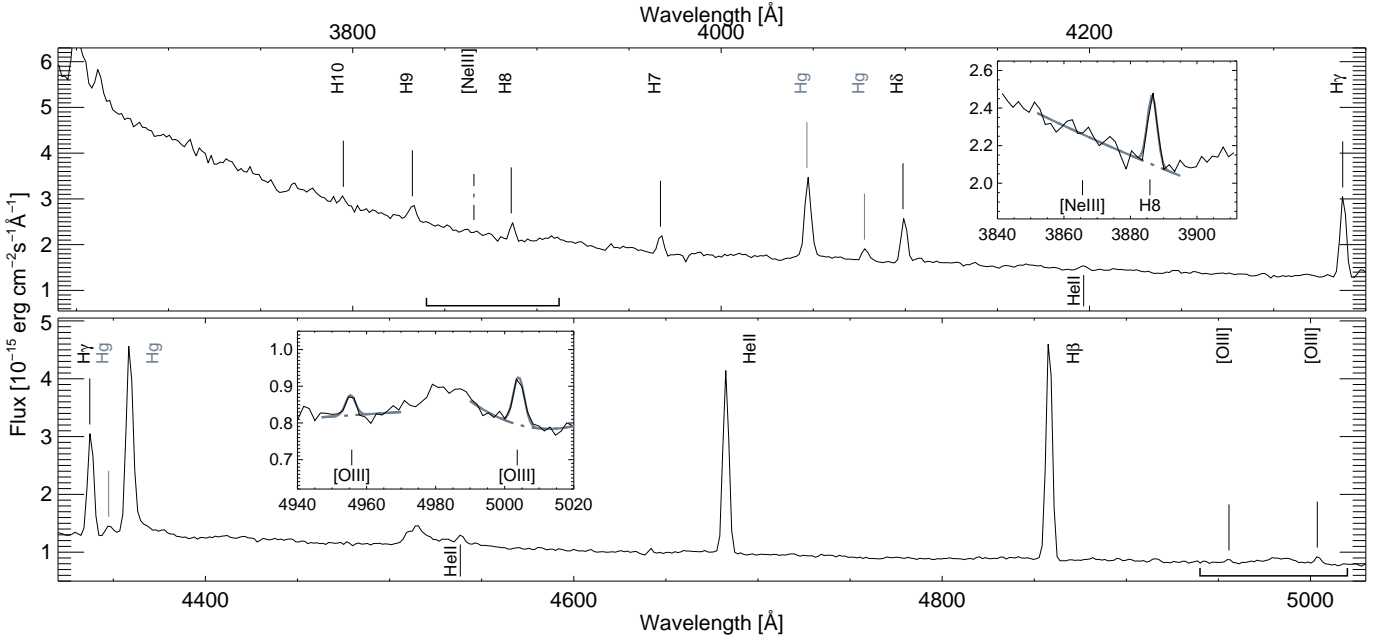}
\caption{The co-added blue spectrum of {\obj} using 123 spatial elements, including the sky. The upper and lower panels show the wavelength ranges 3640--4350\,{\AA} and 4320--5030\,{\AA}, respectively. The insets in the upper and lower panels show a close-up of the spectrum where H8 and a potential $[\ion{Ne}{iii}]\,\lambda3869$, and $[\ion{O}{iii}]\,\lambda\lambda4959,\,5007$ are found, respectively. Gray thick lines in the inset panels show the line fits. The wavelength ranges of the inset panels are indicated in the spectrum with horizontal lines. The positions of four emission lines of telluric origin are also indicated (Hg). The diffuse emission features at, e.g., $\lambda\!\simeq\!4510\,${\AA} and $\lambda\!\simeq\!4980\,${\AA} are also of telluric origin. For further details see Sect.~\ref{sec:results}.}
\label{sandinf1}
\end{figure*}

We made all line fits using the IFU analysis package \textsc{ifsfit} (Sandin, in prep.) that was initially developed for \citet{SaScRo.:08}. We used a second order polynomial to fit the continuum and Gaussian curves to fit emission lines. Because of less favorable weather conditions in the second night we only used the co-added spectrum of the two exposures of the first night, totaling an exposure time of 5400\,s, see Fig.~\ref{sandinf1}.

The intensities of all object lines we measured are presented in Table~\ref{sandint1}, together with the values of {\rR}, {\rJ}, and {\rT}. We present both raw values, and values where we subtracted the contribution of \ion{He}{ii} Pickering lines from the Balmer lines. Note that errors of our intensity measurements decrease for redder emission lines, reflecting the better signal-to-noise of the spectrum in the redder region. Since the first night was not photometric our value of the integrated flux of {\hb} is very likely somewhat uncertain. Nevertheless, we calculated a spatially integrated flux of {\hb}, using the spectrum where the CS is included, of $1.92\!\times\!10^{-14}\,${\Fl} (Table~\ref{sandint1}), and this value compares well with values given in the literature. {\rR} find values ranging from $1.0\,\ldots\,2.6\!\times\!10^{-14}$\,{\Fl}, and {\rJ} evaluate a total {\hb} flux of $1.5\!\times\!10^{-14}$\,{\Fl}. While {\rJ} correct their flux estimate for the region that is not covered by the slit, the varying values of {\rR} seem to be a result of the width of the slit they use.

We did not detect $[\ion{Ne}{iii}]\,\lambda3869$ (cf.\@ the upper inset of Fig.~\ref{sandinf1}). Since the signal-to-noise of our spectrum appears to be better than in previous studies, with five new measured lines in the nebula (see Table~\ref{sandint1}), it is questionable if anyone has detected $[\ion{Ne}{iii}]\,\lambda3869$.

In order to calculate an upper detection limit of emission lines in our data we proceeded as follows. For a certain wavelength we assumed that we can detect a line with the maximum flux of $\sigma$ above the continuum, where $\sigma$ is the error of the measured flux. We calculated the intensity $I$ of this limiting line by integrating over a triangle of width 7{\AA}, and $I\!=\!7/2\times\sigma$. At the three wavelengths $\lambda\!=\!3790,\,3860,\,\text{and}\,5010\,${\AA} we measured $\sigma\!=\!6.17,\,4.70,\,\text{and}\,1.46\!\times\!10^{-17}\,$\FlA. Dividing the resulting intensities with the intensity of {\hb} we got the following limiting line ratios ($\times\!100$): 1.1 ($\lambda\!=\!3790\,\AA$), 0.85 ($\lambda\!=\!3860\,\AA$), and 0.27 ($\lambda\!=\!5010\,\AA$). Hence, we consider $0.01\hb$ as an upper limit of the line strength of $[\ion{Ne}{iii}]\,\lambda3869$.

\begin{table*}[t]
\caption{Flux measurements of {\obj}}
\label{sandint1}
\begin{tabular}{lcr@{\hspace{5pt}}lr@{\hspace{5pt}}lr@{\hspace{5pt}}lr@{\hspace{5pt}}lr@{\hspace{5pt}}lrr}
\hline\hline
\noalign{\smallskip}
Emission line          & $\lambda_0$ [\AA] &
      \multicolumn{2}{c}{PMAS$_{\text{raw}}$} & \multicolumn{2}{c}{PMAS$_{\text{corr.}}$} &
      \multicolumn{2}{c}{{\rR} (SPM1)} & \multicolumn{2}{c}{{\rR} (CFHT)} &
      \multicolumn{2}{c}{{\rJ}$^{\text{a}}$} & \multicolumn{1}{c}{{\rT}} & $I(\text{case B})$\\[1.5pt]
\hline
\noalign{\smallskip}
$[\ion{O}{iii}]$  & 5006.84 &   3.36 & (0.33) &   3.48 & (0.35) &   2.7 & (1.3)  & 2.87   &(0.85) &  3 & (1) & 3.1 &       \\
$[\ion{O}{iii}]$  & 4958.92 &   1.27 & (0.29) &   1.33 & (0.30) &       &        &        &       &    &     &     &       \\
{\hb}             & 4861.32 & 100.00 & (0.49) & 100.00 & (0.49) & 100.0 & (2.1)  & 100.0\xx  & (1.7) & 100&     & 100 & 100.0 \\
$\ion{He}{ii}$    & 4685.65 &  78.72 & (0.62) &  82.11 & (0.64) &  76.1 & (2.3)  &  78.6\xx  & (1.5) & 77 & (3) &  92 &       \\
$\ion{He}{ii}$    & 4541.59 &   2.96 & (0.39) &   3.09 & (0.40)  \\       
H$\gamma$         & 4340.45 &  45.49 & (0.60) &  45.40 & (0.63) &  41.0 & (1.6)  &  42.05 & (0.98)& 39 & (3) &  56 &  47.6 \\
$\ion{He}{ii}$    & 4199.83 &   2.48 & (0.59) &   2.59 & (0.61)  \\
H$\delta$         & 4101.74 &  23.81 & (0.64) &  23.69 & (0.66) &  21.4 & (2.8)  &  20.6\xx  & (1.2) & 17 & (2) &  30 &  26.6 \\
H7                & 3970.07 &   9.70 & (0.73) &  10.11 & (0.76) &  10.0 & (2.5)  &   6.11 & (0.64)&  4 & (1) &  11 &  16.4 \\
H8                & 3889.06 &   8.63 & (0.89) &   9.00 & (0.93) &       &        &   2.92 & (0.73)&    &     &     &  10.8 \\
$[\ion{Ne}{iii}]$ & 3868.80 & $<\!1$ &        & $<\!1$ &        &       &        &   1.04 & (0.52)& 15 & (7) &     &       \\
H9                & 3835.40 &   8.22 & (0.99) &   8.6\xx  & (1.0)  &       &        &        &       &    &     &     &   7.5 \\
H10               & 3797.91 &   3.4\xx & (1.2)  &   3.3\xx  & (1.2)  &       &        &        &       &    &     &     &   5.8 \\
$F$(\hb)$^\text{b}$      && 1.924  & (0.009)&        &        & 1.82  &(0.03)  &   2.55 & (0.03)&\multicolumn{2}{c}{1.47}     & 1.19 \\
\hline
\noalign{\smallskip}
\multicolumn{14}{l}{$^{\text{a}}$ The line ratios are corrected for an extinction value of $E(B-V)\!=\!0.03$ and for a $\sim\!5\%$ contribution of $\ion{He}{ii}$-lines to Balmer lines.}\\
\multicolumn{14}{l}{$^{\text{b}}$ This is the total flux in emission measured for {\hb} in units of $10^{-14}$\,\Fl, systematic errors are not considered in the error estimate.}\\
\end{tabular}\\
\textsc{Note.}--- Errors are given in parentheses. The emission line names and the rest wavelength $\lambda_0$, and Case B line ratios for the Balmer lines to {\hb}, are given in Cols.~1, 2, and 9. We present our raw values in Col.~3 and values corrected for the contribution of $\ion{He}{ii}$ Pickering lines to the Balmer lines in Col.~4. Columns 5--8 give the values presented by {\rR} (SPM1), {\rR} (CFHT), {\rJ}, and \rT, respectively.
\end{table*}

\section{Data analysis using hydrodynamic models}\label{sec:dataanalysis}
Consequences for estimates of abundances in the context of photoionization models and the assumption of thermal equilibrium are poorly known. In a study based on hydrodynamical models using time-dependent ionization \citet[hereafter {\rP}; also see \citealt{ScJaSt2:05}]{ScJaSa:09} demonstrate that metal-poor nebulae with low densities are prone to deviations from thermal equilibrium, because heating by photoionization is no more balanced by \emph{line} cooling only. In extreme cases the electron temperature is also controlled by \emph{expansion} cooling, which results in \emph{lower} electron temperatures compared to the (standard) equilibrium case by up to 30\%, although ionization is still close to equilibrium. Differences are insignificant when solar abundances are used instead \citep{PeKiSc.:98}.

In this section we combine our observed line strengths and UV-data obtained with HST-STIS (Jacoby, priv.\ comm.) with outcome of our time-dependent hydrodynamic models in order to estimate abundances. The UV-spectrum we used is publicly available and can be retrieved from the HST archive (see proposal 9466, PI: Garnavich); it is also published by {\rJs} (see Fig.~1 therein).

 At first we present our modeling approach in Sect.~\ref{sec:discphys}, thereafter our model sample in Sect.~\ref{sec:discmodels} and a discussion of how we determine our abundances in Sect.~\ref{sec:discabundances}.

\subsection{Physical properties of our time-dependent models}\label{sec:discphys}
Our one-dimensional radiation hydrodynamic (RHD) models of envelopes of PNe are described in detail in \citet[also see references therein]{PeKiSc.:98,PeScSt.:04}. We emphasize that our models calculate ionization, recombination, heating, and cooling time-dependently at every time step. The cooling function is composed of the contribution of all considered ions, and for every individual ion up to 12 ionization stages are taken into account. Physical input parameters to the models include properties of the coupled CS model, element abundances, and the density and velocity structures of matter in the envelope.

In calculating the models the wind of the CS was used as input at the inner boundary of the grid. The full model evolution was thereafter followed across the Hertzsprung-Russell diagram for about 15\,000 years until (partly) recombination sets in close to the turn-around point (at an age of about 10\,000\,yr), and into subsequent stages of re-ionization due to advanced expansion. An important feature of our models is that we, at any time, can switch off all time-dependent terms and simultaneously fix the density structure and the radiation field. The models are thereafter evolved until they settle into equilibrium, we are then able to study differences between dynamical and static models (also see {\rP}). We refer to these models as equilibrium models, in comparison to the dynamical models. We calculated surface brightnesses, emission line profiles and strengths of individual lines using a supplementary code that is based on a version of \citet{GeAcSz:96}.

\subsection{Properties of our model sample}\label{sec:discmodels}
A definite mass estimate for the ionizing source of {\obj} is still lacking. {\rTo} identify the nucleus of {\obj} as a close binary, which hampers their analysis, using non-LTE model atmosphere spectra, to derive the mass of the ionizing star. They fit the photospheric Balmer lines of selected optical spectra to obtain the surface gravity $\log g$. Assuming an effective temperature of $\teff\!=\!120\,000\,$K, and that the companion does not contribute to the optical emission, a comparison with stellar evolutionary tracks indicate a mass of $0.88\,M_{\odot}$ that they infer as unrealistically high.

{\rTo} support the final estimate of 0.55--$0.57\,M_{\odot}$ considering only the Population II characteristics of the object (not being in the Galactic plane, having a high radial velocity and being metal-poor), and its relatively high kinematic age that the authors consider a rough estimate for the post-AGB age. In order to be an evolved CS, which still is in a pre-white dwarf stage, a kinematic age of $t_\text{kin}\!=\!16\,000\,$yr demanded a mass range that low (see Fig.~9 in \rTo). The claim of \citet{ToToNa.:07} that the binary core of {\obj} consists of one component of $0.565\,\Msun$ and $\teff\!\simeq\!58\,000\,$K, and a second component of $0.85\,\Msun$ and $\teff\!\simeq\!170\,000\,$K, agrees poorly with their previous results. For instance, the hydrogen and helium lines will hardly be fitted by an object where $\teff\!=\!58\,000\,$K. The ionization of the nebula is likewise inconsistent with an ionizing source where $\teff\!=\!170\,000\,$K, as {\rPT} show.

{\rPT} set {\teff}, the object distance and luminosity, and select models using a range of masses (0.583--0.600$M_{\odot}$) based also on arguments using a kinematic age. Conclusions, which are based on kinematic ages ($R_\text{xxx}/V_\text{yyy}$), are questionable when they are used as evolutionary ages. Errors emerge from uncertain distances, the confusion of matter (Doppler) velocities with structure (shock) velocities, inappropriate (i.e.\ unrelated) combinations of $R_\text{xxx}$ and $V_\text{yyy}$, and neglecting the expansion history of the object \citep[also see Fig.~33 in \rP]{ScJaSt3.:05}. The subscripts (xxx and yyy) denote that various combinations of $R$ and $V$ could be used in order to obtain a kinematic age. For $R$ one could use the outermost radius ($R_\text{out}$), the rim radius, or the radius of any other substructure. For $V$ one could, moreover, use corresponding differential changes ($\dot{R}_\text{xxx}$, although this quantity is rarely available), or alternatively spectroscopic velocities such as the half-width half-maximum (HWHM), HW10\%M of spatially unresolved profiles, or any velocity component of a decomposition of a spectrum of a spatially resolved profile.

Because the properties and evolutionary history of {\obj} are uncertain we restricted our analysis to one stellar evolutionary track as input for our RHD models, and used a post-AGB (CS) model of $0.595\,M_{\odot}$. Using a similar approach as {\rTo} we only considered one stellar component when evolving the nebula, and assumed that the influence from the companion is weak. The stellar wind from the CS is calculated according to \citet{MaSc:91}. Its dependence on the metallicity (in this case C, N, and O) is approximately accounted for by correction factors; we used $\dot{M}\!\propto\!Z^{0.69}$ \citep{ViKoLa:01} and $v_\infty\!\propto\!Z^{0.13}$ \citep{LeRoDr:92}, whereby $L_\text{wind}\!=\!0.5\dot{M}v_\infty^2\!\propto\!Z^{0.95}$. The CS radiates as a black body. For the AGB wind we, moreover, adopted a power-law density profile $\rho\!\propto\!r^{-\alpha}$ ($\alpha\!=\!3$--3.25, cf.\ Sect.~\ref{sec:discabundances}) and a constant outflow velocity $v\!=\!10\kms$ \citep[cf.][hereafter \rPtwo]{ScJaSt.:05}. We used the radial domain $4.0\times\!10^{14}\!\le\!r\!\le\!2.8\!\times\!10^{18}\,\text{cm}$. The models are normalized such that $n\!=\!10^5\,\text{cm}^{-3}$ at $r\!=\!3\times10^{16}\,$cm. The abundances are based on scaled values of mean Galactic disk PNe abundances ($Z\ugd$; this abundance distribution is first quoted by \citealt{PeKiSc.:98}) for nine elements, see Table~\ref{sandint2}; except for carbon and nitrogen $Z_{\text{GD}}$ is close to solar. The abundance values $\epsilon_{i}$ are specified in (logarithmic) number fractions relative to hydrogen, i.e.\ $\epsilon_{i}\!=\!\log\,(n_{i}/n_{\text{H}})\!+\!12$. Abundance distributions, which are similar to $Z\ugd$, are used by for example, \citet{Pe:91}, \citet{KiBa:94}, \citet{ExBaWa:04}, and \citet{HyPoFe:04}. We summarize all model parameters and properties in Table~\ref{sandint4}. Selecting metallicities we made use of the set of six models of {\rP} that we refined with four additional models (these additional models are marked with the prefix $^\star$): $3Z\ugd$, $Z\ugd$, $Z\ugd/3$, $Z\ugd/10$, $^{\star}Z\ugd/15$, $^{\star}Z\ugd/20$, $^{\star}Z\ugd/25$, $Z\ugd/30$, $^{\star}Z\ugd/60$, and $Z\ugd/100$.

\begin{table}[t]
\caption{Mean Galactic disk element abundance distribution ($Z_{\text{GD}}$)}
\tabcolsep=4.2pt
\begin{tabular}{lcccccccccc}
\hline\hline\noalign{\smallskip}
                 & H & He    & C    & \multicolumn{1}{c}{N} & O    & Ne   &  S     &  Cl    &  Ar  \\
\hline\noalign{\smallskip}
$Z\ugd$          & 12.00 & 11.04 & 8.89 & 8.39 & 8.65 & 8.01 &  7.04  &  5.32  &  6.46\\
\hline
\end{tabular}
\label{sandint2}
\end{table}

\begin{figure}[t]
\centering
\includegraphics[width=8.8cm]{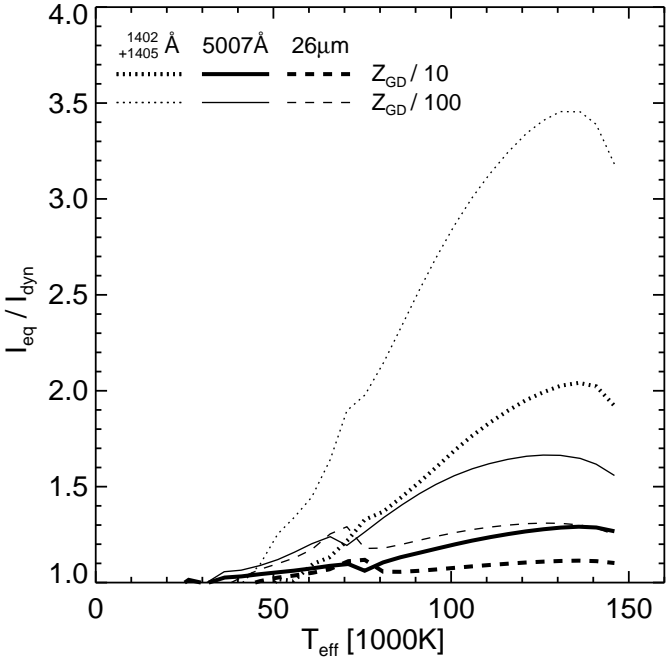}
\caption{Comparison between emission line strengths of dynamical (dyn) and equilibrium (eq) nebular models around a $0.595\,\Msun$ central star evolving across the H-R diagram. The two shown model sequences use abundances $Z\ugd/10$ (thick lines) and $Z\ugd/100$ (thin lines). We show the dyn/eq line strength ratios (I$_\text{eq}/$I$_\text{dyn}$) of three collisionally excited oxygen lines, viz.\@ $\ion{O}{iv}]\,\lambda\lambda1402\!+\!1405$ (dotted lines), $[\ion{O}{iii}]\,\lambda5007$ (solid lines), and $[\ion{O}{iv}]\,\lambda26\,\mu$m (dashed lines), as a function of \teff. The evolution is only traced until maximum $\teff$\ is reached in order to avoid confusion. See Fig.~\ref{sandinf9} for absolute values on the intensities. For further details see Sect.~\ref{sec:discmodels}.}
\label{sandinf2}
\end{figure}

In order to illustrate differences between dynamical and equilibrium models we present line strength ratios of three collisionally excited oxygen lines in the UV, optical, and infrared wavelength ranges in Fig.~\ref{sandinf2}. Representing the model evolution the line ratios are shown as a function of the CS effective temperature \teff. The figure shows that lines of dynamical models are weaker than those of equilibrium models for $\teff\ga50\,000\,$K. The reason is that dynamical models have a lower electron temperature (cf.\ Sect.~3.1.3 and Fig.~16, both in {\rP}). We do not consider young PNe, where $\teff\!<\!50\,000\,$K; in such objects the ionization equilibrium is disturbed for a short time, with the consequence that the ionization is somewhat overestimated in the equilibrium models.

Discrepancies are larger with a lower metallicity, compare the thin lines with the thick lines in Fig.~\ref{sandinf2}, and shorter wavelengths, compare the dotted lines with the dashed lines. When we compare the line intensities at $\teff\!\simeq\!130\,000\,$K we see that, as expected, the difference is the largest for the UV-line $\ion{O}{iv}]\,\lambda\lambda1402\!+\!1405$ that is up to 250\% stronger in the equilibrium case of the $Z\ugd/100$ sequence. The optical line $[\ion{O}{iii}]\,\lambda5007$ is stronger by about 70\%. The infrared line is the least affected, it is up to about 30\% stronger in thermal equilibrium.

Because differences in line strengths depend strongly on the input density and the velocity structure, these values should only be regarded as indicative. Nevertheless, non-equilibrium effects introduce an ambiguity to the determination of electron temperatures, hence elemental abundances, which is neglected when using standard photoionization codes (cf.\ Sect.~\ref{sec:discussion}).

\subsection{Estimating abundances for \obj}\label{sec:discabundances}
A full abundance analysis of {\obj} is beyond the scope of this work, but as an example we determine abundance estimates that are based on our hydrodynamical models. Our model grid is not large enough to allow iteration of all parameter dimensions. The primary goal of this study is instead to find a model that agrees reasonably with the observational quantities, and then use this \emph{best-match} model to elaborate on the influence of non-equilibrium effects.

We used the following four criteria to find such a best-match model: emission line strengths should match observed values, the model \ha~emission line profile should match the observed line profile \citep[such as Fig.~1 of][]{RiLoSt.:03}, the model \ha~surface-brightness distribution should match the observed distribution (such as Fig.~3 of \rR), and both distance estimates from the \hb~flux of the object and from the corresponding apparent size, which is obtained from the surface-brightness distribution, should be in fair agreement. In addition to these properties that apply to the nebula, the visual magnitude of the central star is also indicative of the distance. However, since we have only used one single stellar track (with a stellar mass $M\!=\!0.595\,M_{\sun}$) such a visual magnitude is difficult to match properly. The emission lines and their line strengths that we used in our study, are shown in Cols.~1--3 of Table~\ref{sandint3}.

\begin{table}[t]
\caption{Observational vs.\ modeled line strengths}
\label{sandint3}
\tabcolsep=5.5pt
\begin{tabular}{llr@{\hspace{5pt}}lrrc}
\hline\hline
\noalign{\smallskip}
ID     & $\lambda_0$ [\AA] & \multicolumn{2}{c}{observed}  & \multicolumn{2}{c}{model} & ratio\\[1.5pt]
       &                   &              & & dyn & \multicolumn{1}{c}{eq} & eq/dyn\\[1.5pt]
\hline
\noalign{\smallskip}
 $\ion{N}{v}$      & 1238+1242  &     426$^{\text{a}}$  & (40)    & 330 &  697 & 2.11\\
 $\ion{N}{iv}]$    & 1486       &      87$^{\text{a}}$  & (30)    & 124 &  276 & 2.23\\\noalign{\smallskip}
 $\ion{C}{iv}$     & 1548+1550  &     660$^{\text{a}}$  & (50)    & 668 & 1319 & 1.97\\
 $\ion{C}{iii}]$   & 1906+1909  &      55$^{\text{a}}$  & (30)    &  31 &   64 & 2.06\\\noalign{\smallskip}
 $\ion{O}{iv}]$    & 1402+1405  & $<\!37$$^{\text{a}}$  &         &   6 &   13 & 2.17\\
$[\ion{O}{iii}]$   & 5007       &     3.5$^{\text{b}}$  & (0.4)   & 3.3 &  4.4 & 1.33\\
$[\ion{O}{iv}]$    & 26\,$\mu$m &      --               &         &  26 &   30 & 1.15\\\noalign{\smallskip}
$[\ion{Ne}{iv}]$   & 2422+2425  &      25$^{\text{a}}$  & (12)    &  38 &   62 & 1.63\\
$[\ion{Ne}{v}]$    & 3426       &      86$^{\text{a}}$  & (9)     &  78 &  119 & 1.52\\
$[\ion{Ne}{iii}]$  & 3869       &  $<\!1$$^{\text{b}}$  &         & 0.4 &  0.5 & 1.25\\
$[\ion{Ne}{v}]$    & 14\,$\mu$m &      --               &         &  90 &  102 & 1.13\\\noalign{\smallskip}
$\ion{He}{ii}$     & 4686       &      82$^{\text{b}}$  & (1)     &  82 &   80 & 0.98\\
\hline
\noalign{\smallskip}
\multicolumn{7}{l}{$^{\text{a}}$ Jacoby (priv.~comm.), $^{\text{b}}$ this paper}
\end{tabular}
\textsc{Note.}--- All values are specified in units of $\hb\!=\!100$. Columns 1 \& 2 give the emission line names and the rest wavelength $\lambda_0$. Measured values of {\obj} are given in Col.~3, with errors in parentheses. Corresponding values of dynamic (dyn) and equilibrium (eq) models, and their ratio, are given in Cols.~4--6 (cf.\ Sect.~\ref{sec:discabundances}).
\end{table}

\begin{figure}[t]
\centering\includegraphics[width=8.8cm]{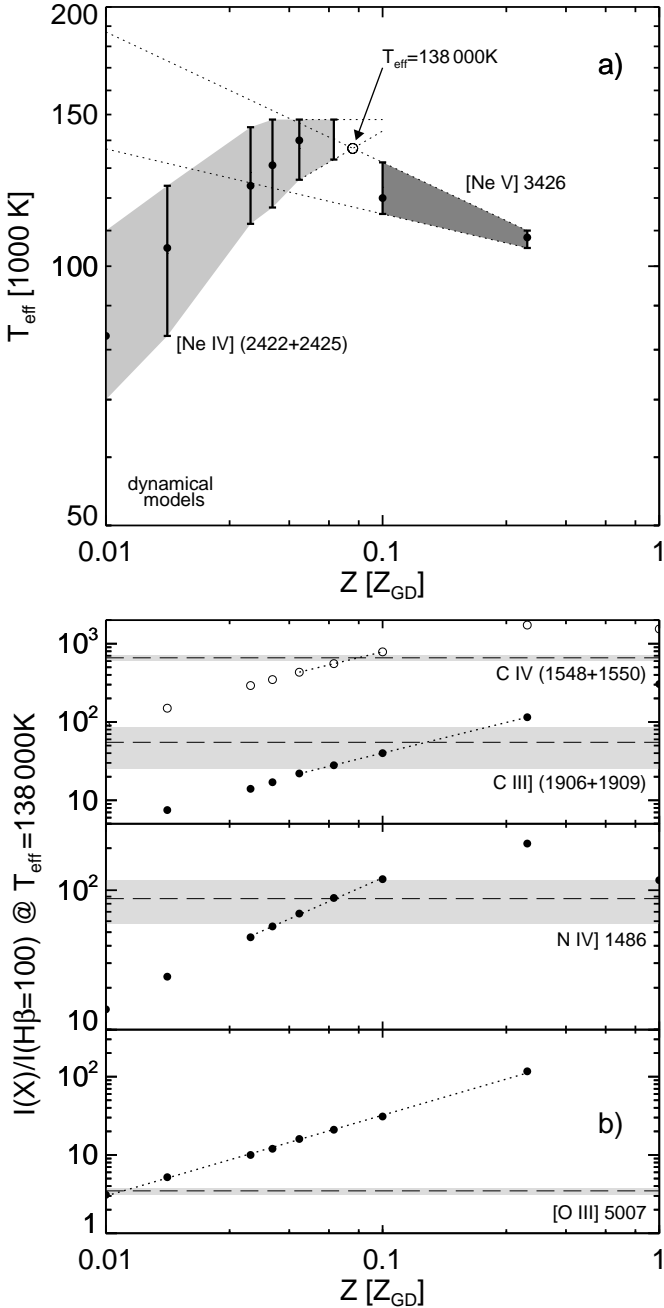}
\caption{For each model sequence panel \textbf{a)} shows the range of {\teff} that corresponds to the observed line strengths of $[\ion{Ne}{iv}]\,\lambda\lambda2422,\,2425$ and $[\ion{Ne}{v}]\,\lambda3426$, compare the plotted values with the neon line strengths in Fig.~\ref{sandinf10}. The value of the effective temperature of our first custom model is indicated with an open circle ($\circ$; $\teff\!=\!138\,000\,$K). In panel \textbf{b)} each sub-panel shows the line strength of all models (but 3Z) that are all evaluated at $\teff\!=\!138\,000\,$K, for four different emission lines. The horizontal dashed lines indicate the observed values. In both panels error intervals are indicated with gray regions. All axes are logarithmic. For further details see Sect.~\ref{sec:discabundances}.}
\label{sandinf3}
\end{figure}

Since we have not been able to measure $[\ion{Ne}{iii}]\,\lambda\,3869$ we cannot determine {\teff} using that line. Instead we proceeded as follows. Using the existing measurements of the two neon lines $[\ion{Ne}{iv}]\,\lambda\lambda\,2422\!+\!2425$ and $[\ion{Ne}{v}]\,\lambda\,3426$ we determined a range of effective temperatures where these measurements match our model emission line strengths (see Fig.~\ref{sandinf10}). The result is shown in Fig.~\ref{sandinf3}a. The upper limit of the temperature range corresponding to the $[\ion{Ne}{iv}]$-line is set by the maximum stellar temperature possible, in this case $\max(\teff)\!=\!146\,870\,$K. Note that the extent of the two regions of the effective temperatures of $[\ion{Ne}{iv}]$ and $[\ion{Ne}{v}]$ that do not overlap, depends on both observations and model initial conditions. We extrapolated both regions in order to find the nearest point of intersection ($Z_\text{Ne}$, \teff) that we then used as two of the initial values when iterating our models. From this procedure we found $\teff\!=\!138\,000\,$K.

For carbon, nitrogen and oxygen we evaluated line strengths at {\teff} for all model sequences (Figs.~\ref{sandinf7}--\ref{sandinf9}); the variation of model line strengths with stellar effective temperature is moderate at $\teff\!=\!138\,000\,$K (see below). We show the results in Fig.~\ref{sandinf3}b, together with power law fits in relevant regions. By a comparison with the observational data we then found a first set of object abundances. For sulfur, chlorine and argon we simply reduced the respective mean Galactic disk value using the mean scaling value that we found for carbon, nitrogen, oxygen, and neon.

In a final step, during the creation of our first custom model, we adjusted the helium abundance. Figure~\ref{sandinf11} shows that model line strengths of helium are nearly independent of metal abundances. The same figure also shows that the variation of each model sequence with the effective temperature is small for evolved objects (where $\teff\!>\!10^5\,$K). We reduced the initial helium abundance by one third, as is suggested by the ratio of model-to-observed line strength. With this set of initial abundances we then calculated additional custom models with modified abundances in order to converge on the observed line strengths. In this process we iterated element abundances, which were the closest to the observed values, accounting for the size of the error bars of the different lines. Setting the carbon abundance we first gave $\ion{C}{iv}\,\lambda\lambda1548\!+\!1550$ a stronger weight than $\ion{C}{iii}]\,\lambda\lambda1906\!+\!1909$, since its error is smaller (also see Fig.~\ref{sandinf3}b). 

In the course of our iterative calculations we found that the model {\hb} flux did not match the apparent size of the object at a common distance. In order to achieve a less extended model we modified the power law density distribution of the AGB wind by replacing $\alpha\!=\!3$, that is used in all other models of this study, with $\alpha\!=\!3.25$; keeping the density normalization as before (see Sect.~\ref{sec:discphys}). By this approach changes to resulting model emission line strengths should be small. In {\rPtwo} we show that a steeper density gradient results in faster expansion rates of the leading shock of the shell, i.e.\ the outer edge of the PN. Therefore our new choice of $\alpha$ may appear counterintuitive when forming a geometrically smaller PN. The size of a PN, however, depends on the expansion velocity integrated over the entire expansion period. Early on a high circumstellar density at the inner edge of the model domain (due to our choice of a power law) prevents nebular matter from becoming rapidly ionized. The formation of a D-type ionization front is thereby delayed as the acceleration starts later with a steeper density gradient, and the overall PN lifetime thereby becomes shorter. In the case of $\alpha\!=\!3.25$ a higher expansion velocity cannot compensate for the simultaneous shortening of the expansion period.

\begin{figure*}[t]
\centering
\includegraphics[width=17.8cm]{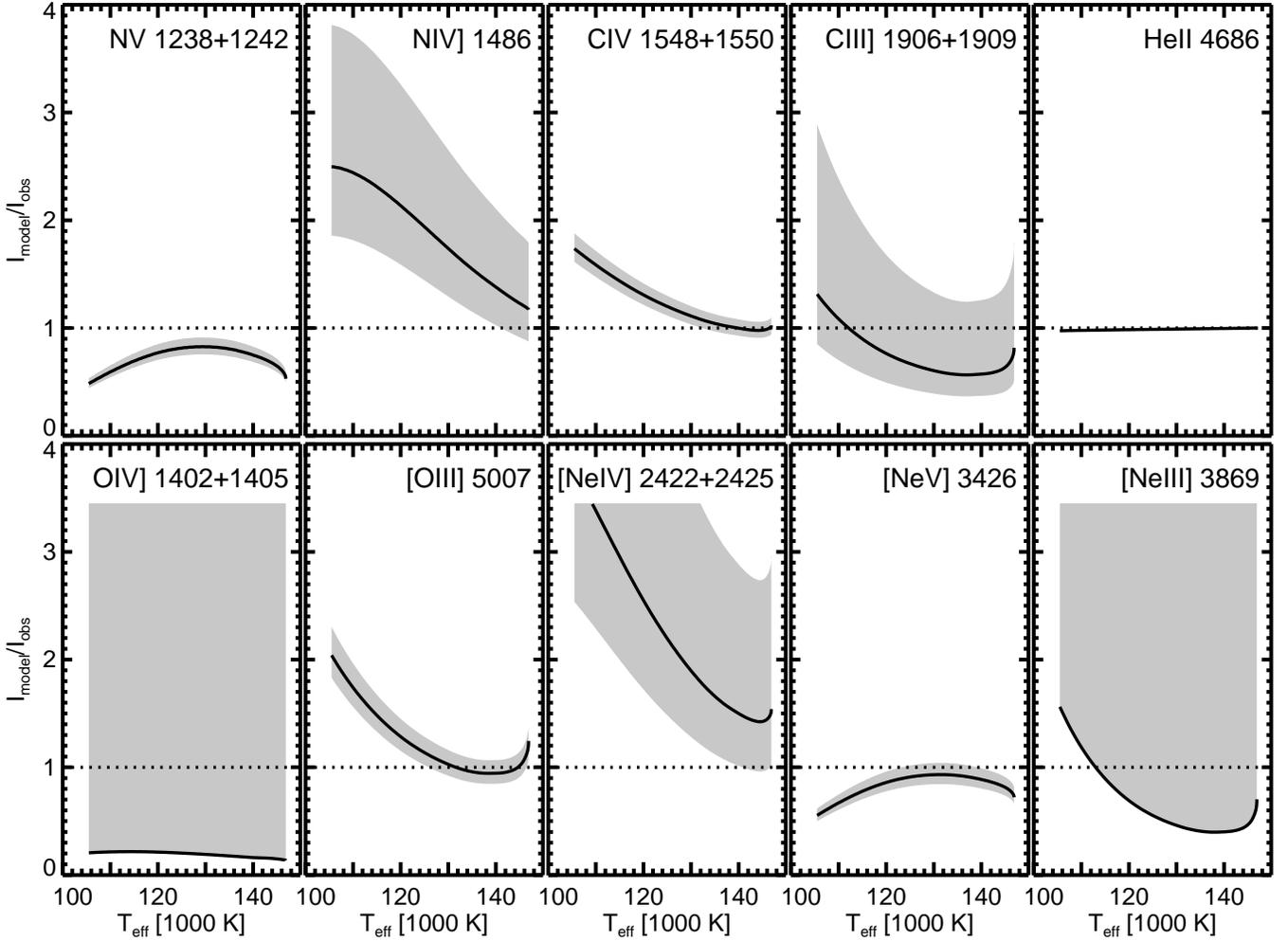}
\caption{This figure shows the evolution of emission line strengths of our best-match model, as a function of the effective temperature, \teff. Using the ten observed emission lines of Table~\ref{sandint3} we show model-to-observed line ratios (solid lines). The gray regions mark the corresponding error bars. Note that only upper limits were observed for $\ion{O}{iv}]\,\lambda\lambda1402,\,1405$ and $[\ion{Ne}{iii}]\,\lambda3869$. For further details see Sect.~\ref{sec:discabundances}.}
\label{sandinf4}
\end{figure*}

\begin{table}[t]
\caption{Summary of our best-match model parameters}\label{sandint4}
\tabcolsep=4.2pt
\begin{tabular}{lll}
\hline\hline\noalign{\smallskip}
Parameter & \obj & \multicolumn{1}{c}{$Z\ugd$}\\
\hline\noalign{\smallskip}
Spectrum                 & black body\\
Stellar mass, $M$        & 0.595\,\Msun\\
Model age, $t$           & $8982\,$yr\\
Stellar effective temperature, \teff & 138\,049\,K\\
Stellar luminosity, $L$  & $2994\,\Lsun$\\
Central star wind, $L_\text{wind}=0.5\dot{M}v^2_\infty$ & \multicolumn{1}{c}{$\dot{M}\propto Z^{0.69}$, $v_\infty\propto Z^{0.13}$}\\
AGB wind                 & \multicolumn{2}{l}{$\rho\propto r^{-3.25}$, $v\!=\!10\,\kms$,}\\
                         & \multicolumn{2}{l}{$n_{r=3\times10^{16}\,\text{cm}}\!=\!1\!\times\!10^5\,\ccm$}\\[1.0ex]
\hline\\[-1.5ex]
Abundances, $\epsilon_i\!=\!\log\,(n_i/n_\text{H})+12$:\\
\quad He & 10.88 & 11.04\\
\quad C  & \xx7.90  & \xx8.89 \\
\quad N  & \xx7.47:  & \xx8.39 \\
\quad O  & \xx6.74  & \xx8.65 \\
\quad Ne & \xx6.96  & \xx8.01 \\
\quad S  & [5.94]   & \xx7.04 \\
\quad Cl & [4.22]   & \xx5.32 \\
\quad Ar & [5.36]   & \xx6.46 \\[1.0ex]
\hline\\[-1.5ex]
Distance, $d$            & 18\,kpc\\
Visual magnitude, $m_\text{V}$ & 19.5\,mag\\
Nebular density, $\langle\Ne\rangle$ & 65\,\ccm\\
Nebular temperature, $\langle\Te\rangle$ & 21\,100\,K\\
Nebular \hb-luminosity, $L(\hb)$ & 0.193\,\Lsun\\
Model HWHM velocity, $V_\text{HWHM}$ & $41.8\,\kms$\\[1.0ex]
\hline
\noalign{\smallskip}
\end{tabular}
\textsc{Note.}--- The element abundances $\epsilon_{i}$ are used as input in the calculation of our radiation hydrodynamic models; the values of S, Cl, and Ar are not fitted, but only scaled. $Z\ugd$ denotes the mean abundance distribution in the Galactic disk (cf.\ Sect.~\ref{sec:discphys}).
\end{table}

The abundance distribution of our best-match model, and all relevant model properties, is given in Table~\ref{sandint4}, and resulting emission line strengths are given in Col.~4 of Table~\ref{sandint3}. We also show model-to-observed line strength ratios in Fig.~\ref{sandinf4}. This figure illustrates a weak dependence with effective temperature at values about $\teff\!\simeq\!138\,000\,$K. Since most lines depend only weakly temperatures about this value, the precise value of {\teff}, for say $138\,000\!\pm\!5000\,$K, is uncritical to the ionization structure. We could not achieve a simultaneous agreement for the two nitrogen lines. The high value of 426 for $\ion{N}{v}\,\lambda\lambda1238\!+\!1242$ could not be reached with any of our models (Fig.~\ref{sandinf8}), which is why the nitrogen abundance of our best-match model should be considered approximate. $\ion{O}{iv}]\,\lambda\lambda1402\!+\!1405$ can, furthermore, hardly be identified in the spectrum of {\rJs}. We consider the value of 37 a very conservative upper limit -- compare with the value of $87\pm30$ for $\ion{N}{iv}]\,\lambda1486$. A possible blending with $\ion{Si}{iv}\,\lambda\lambda1394\!+\!1403$ should not be excluded.

\begin{figure*}[t]
\centering
\includegraphics[width=17.8cm]{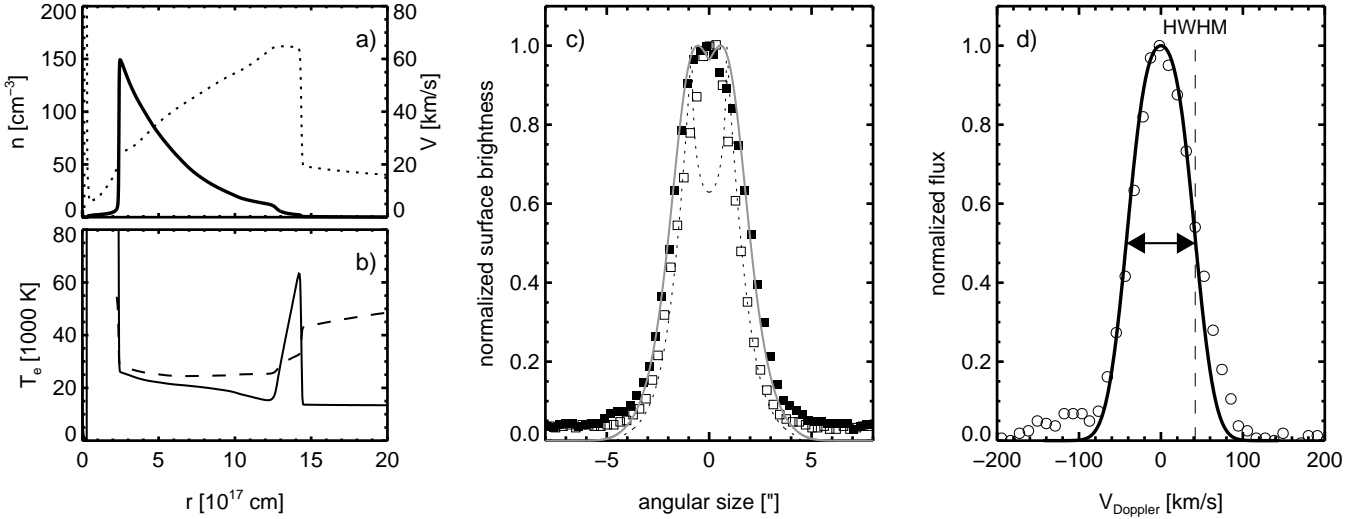}
\caption{This figure shows the basic physical properties of our best fit model at the stellar parameters $L\!=\!2\,985\,\Lsun$ and $\teff\!=\!138\,152\,$K, at an age of $t\!=\!8\,988\,\text{yr}$. The four panels show: \textbf{a)} the radial structure of of the particle density (thick line) and the gas velocity (dotted line), \textbf{b)} radial structures of the electron temperature for the dynamical model (solid line) and the equilibrium model (dashed line), \textbf{c)} a comparison between the \ha surface-brightness distribution of the model at a assumed distance of $18.3\,$kpc and the observational data of {\rR} (filled/open squares = semi-major/semi-minor axis, see text for details) with the actual \ha-image (dotted line) and an image that results with a seeing of $1\farcs5$ (solid line), and \textbf{d)} a comparison between the \ha~emission line profile of the model (solid line) and the observation of \citet[see Fig.~1 and slit 7; open circles $\circ$]{RiLoSt.:03}. The simulated profile was additionally broadened by a Gaussian with FWHM$=\!26\,\kms$, in order to be consistent with their observation. The observed left wing is due to emission of $\ion{He}{ii}\,\lambda6560$. For further details see Sect.~\ref{sec:discabundances}.}
\label{sandinf5}
\end{figure*}

In Fig.~\ref{sandinf5} we show structural and kinematic properties of our best-match model; the CS has evolved to $t\!=\!8982\,$yr, $L\!=\!2994\,\Lsun$, and $\teff\!=\!138\,049\,$K (Table~\ref{sandint4}). The mean electron density is $\langle\Ne\rangle\!=\!65\,\mbox{cm}^{-3}$. The density (Fig.~\ref{sandinf5}a) shows a gradual decline with increasing radius. Neither the density nor the {\ha} surface-brightness structure (Fig.~\ref{sandinf5}c) show a distinct double shell morphology. The same panel compares the observed data of {\rR} (see Fig.~3 therein) with the outcome of our model, using a distance of $d\!=\!18.3\,$kpc. We show the real structure with a central dip that is due to the hot bubble, and the structure under the seeing conditions of the observations ($1\farcs5$). A comparison of our observed {\hb} flux, $F(\hb)\!=\!1.924\times10^{-14}$\,\Fl, and the {\hb}-luminosity of the model, $L(\hb)\!=\!0.193\,\Lsun$, yields a distance of $d\!=\!17.95\,$kpc -- that agrees well with the one of the surface-brightness structure. We adopt $d\!=\!18\,$kpc as a final value of the distance. In our models we use exclusively black bodies for the CSs. Applying the distance estimate of $d\!\approx\!18\,$kpc to the bolometric luminosity of the model, $L\!=\!2\,994\,\Lsun$, yields $m_\text{V}\!=\!19.5\,$mag\footnote{{\rR} measure $m_\text{V}\!=\!17.9\,$mag. Assuming this value instead our best-match model would shift to a corresponding distance of only $d\!=\!8.6\,$kpc. This discrepancy in our third distance estimate can be explained by our selected stellar mass (one single track of 0.595\,\Msun), which should also be iterated in order to achieve a better agreement. Of course, our model cannot reproduce the actual intensity if the nucleus really is a double degenerate.}.

Throughout the nebula the matter velocity gradient is positive with increasing radius (Fig.~\ref{sandinf5}a), reaching a maximum velocity of $v\!\simeq\!65\kms$. In the adjacent (radiative) shock layer at $12.5\!\la\!r\!\la\!14.5\!\times\!10^{17}\,\text{cm}$ the velocity is about constant. The simulated emission line profile (Fig.~\ref{sandinf5}d) resembles the long slit Echelle spectral analysis of \citet[see Fig.~1, slit 7]{RiLoSt.:03}. Our one-dimensional model cannot be fully applied to this slightly non-spherical PN and its asymmetric line profiles. The observed HWHM velocity, $V_\text{HWHM}\!=\!42.5\,\kms$, is, however, well matched by our model with $V_\text{HWHM}\!=\!41.8\,\kms$.

In Fig.~\ref{sandinf5}b we show the radial electron temperature structure. The temperature peak behind the outer shock does not contribute to the mean temperature due to low ion densities in that region. We will in the following subsection discuss the consequences of non-equilibrium conditions for the electron temperature and, consequently, for the strengths of collisionally excited lines. 

\section{Discussion}\label{sec:discussion}
In order to study differences in the outcome of our time-dependent and static (equilibrium) models we should, ideally, also calculate a corresponding best-match equilibrium model by the same procedure we used to find the best-match dynamical model. As such an approach is extremely time-consuming we instead compare our outcome with literature values, which are all based on standard photoionization codes (these correspond to our equilibrium models). We present our abundances anew in Table~\ref{sandint5} together with a compilation of literature values, which are derived using observed emission lines. Errors of individual estimates are specified where such values are provided. Differences between estimates and physical assumptions of different sources are large in general. Since we focus on understanding general trends of values, and not on providing final abundances, we have not estimated errors of our values. This is also difficult to do with our models where abundances are input parameters, and not the outcome.

\begin{table*}[t]
\caption{Literature compilation of abundances estimates for \obj}
\label{sandint5}
\tabcolsep=5pt
\begin{tabular}{rlcccr@{\:}lr@{\:}lr@{\:}lcr@{\:}lr@{\:}lr@{\:}l}
\hline\hline
\noalign{\smallskip}
Ref. & \multicolumn{1}{c}{spectral}& \multicolumn{1}{c}{\teff} & \multicolumn{1}{c}{$\langle\Te\rangle$} & He & \multicolumn{2}{c}{C} & \multicolumn{2}{c}{N} & \multicolumn{2}{c}{O} &  Ne & \multicolumn{2}{c}{C/O} & \multicolumn{2}{c}{N/O} & \multicolumn{2}{c}{Ne/O}\\
 & \multicolumn{1}{c}{domain}& \multicolumn{1}{c}{$[10^3\text{K}]$} & \multicolumn{1}{c}{$[10^3\text{K}]$}\\[1.5pt]
\hline
\noalign{\smallskip}
\rT  & O&150&--&&&&&&6.3\xx&(0.5)\\
\rJ  & O&100&17.6& 10.82 &         &&         &&     6.93 && 7.47&       &&       &&       3.47\\
\rR  & O &100&30& 10.9$\phantom{0}$  &&     && & 6.15 & (0.35) & 6.35  &     &     &&     &0.5$\phantom{0}$&(0.3)\\
\rPT & O&130&30&10.91&    &&   &&$7.5\phantom{0}$&(0.3)&6.65&&&&&0.14&(0.14)\\
\rS  & O+U& -- & \multicolumn{1}{c}{--} &   & 7.51&(0.15) & 6.87 & (0.19) & 6.85 &(0.25) & 6.6$\phantom{0}$ & 4.7$\phantom{0}$&(1.1)& 1.05&(0.15) & 0.65&(0.35)\\
\rJs  & O+U& 130 &30$^{\text{a}}$&10.87& 7.58&&6.94&&7.18&&6.66&2.5$\phantom{0}$&&0.58&&0.30\\
this work & O+U& 138 & 21.1&10.88 & 7.90 && \multicolumn{2}{l}{7.47:}& 6.74 && 6.96 & 14$\phantom{.0}$ && 5.4:$\!\!\phantom{0}$ && 1.7$\phantom{0}$\\[2ex]
$Z\ugd$ & &&&11.04 & 8.89 && 8.39 && 8.65 && 8.01 & 1.74&& 0.55 && 0.23\\
BoBn-1 &&&& 11.05 & 8.85 && 8.00 && 7.83 && 7.72 &10.5&& 1.48 && 0.78\\
\hline
\noalign{\smallskip}
\multicolumn{10}{l}{$^{\text{a}}$ Jacoby (priv.~comm.)}
\end{tabular}
\textsc{Note.}--- The table only includes estimates that are based on observed emission lines. Columns~1--4 specify the source reference, the wavelength range (O -- optical, and U -- UV), the stellar effective temperature (\teff) and the mean electron temperature ($\langle\Te\rangle$) used in the study. Columns~5--9 give element abundances using the same units as in Table~\ref{sandint2}. In Cols.~10--12 we also give abundance ratios relative to oxygen. Uncertainties are, were provided, given in parentheses. A colon indicates an uncertain value of our best-match model. The abundances of our best-match model are given in the row marked \emph{this work}. In the last two rows we, for comparison, give the mean abundance distribution of the Galactic disk ($Z\ugd$; Table~\ref{sandint2}) and the halo-PN {\BoBn} \citep[the values of this object are taken from][]{HoHeMc:97}. For further details see Sect.~\ref{sec:discussion}.
\end{table*}

Previous studies of {\obj} present improvements to different parts of the abundance analysis. {\rPT} provide a thorough model analysis, without making own observations, where they account for several physical issues, which were not addressed previously. Notably, they study differences in models assuming case B vs.\ non-case B photoionization, they use different sets of collisional recombination coefficients, and use a stellar atmosphere model of the CS, in addition to the commonly used black-body model. Due to lack of data they calibrate their models using only observational data in the visual wavelength range, and therefore they cannot calculate precise values for the abundances of carbon and nitrogen. The main conclusion of {\rPT} is that the oxygen abundance of previous studies is too low. {\rS} and {\rJs}, moreover, add UV lines, which are sampled with HST-STIS, to their linelist, and can thereby constrain the abundances of carbon and nitrogen better than {\rPT} ({\rS} also announce IR observations using SPITZER). {\rS} also argue that {\rPT} use too high abundances for carbon and nitrogen, and therefore have to use a higher oxygen abundance than is necessary; {\rS} find a lower value on the oxygen abundance than {\rPT}, which is in better agreement with estimates of previous studies.

In agreement with all previous studies, except for {\rPT}, we found that the oxygen abundance of {\obj} is very low. It is difficult to make a meaningful, more detailed, comparison between our abundances and those of {\rT}, {\rJ}, and {\rR} since we use different stellar effective temperatures (mainly). {\rPT} are the latest authors who base their analysis on only optical emission lines. The very thorough analysis these authors make is found to be of small use as there is a strong disagreement between the predicted UV-line intensities of their best model (that agrees best with model M1, cf.\ Table~3 in {\rPT}) and observed UV line strengths (Table~\ref{sandint3}); the difference is (with the exception of $\ion{C}{iii}]\,\lambda1908$) in every case about a factor two. It is in this context worth mentioning that the electron temperature they adopt is about 9000\,K higher than in our study, resulting in a different ionization structure (see below). {\rJs} use a similar electron temperature as {\rPT} in their photoionization models, which is why their results should be affected to a similar degree. Furthermore, although {\rS} include UV-intensities in their analysis they do not provide any temperatures at all and it is impossible to make a meaningful comparison with their abundances.

Compared to the mean abundances of the Galactic disk our values for {\obj} are 1/1.45 (He), 1/9.8 (C), 1/8.3 (N), 1/81 (O), and 1/11 (Ne). Compared to the total mean metallicity of the Galactic disk our value is 1/13. Our abundance estimates relative to oxygen are in better agreement with the values of another halo PN, BoBn-1. In this case our values of C/O, N/O, and Ne/O are 33--260\% higher, although {\obj} is considerably more depleted of metals.

\begin{figure}[t]
\centering
\includegraphics[width=8.8cm]{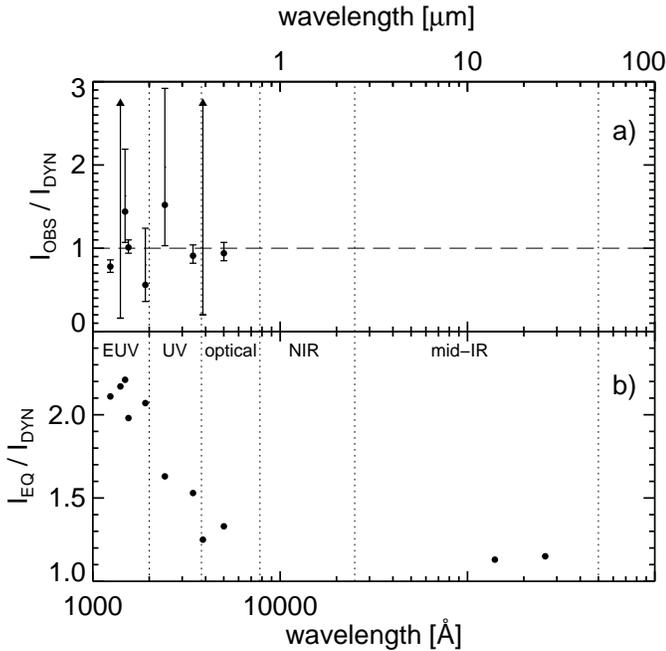}
\caption{In panel \textbf{a)} we show the ratio between the emission line strength values of our best-match model (dyn) and the observed values (cf.\ Table~\ref{sandint3} and Sect.~\ref{sec:discabundances}). For all, but the two nitrogen lines, $\ion{N}{v}\,\lambda\lambda1238\!+\!1242$ and $\ion{N}{iv}]\,\lambda1486$, the model values and observations agree within error bars. In panel \textbf{b)} we show emission line strengths of the thermally relaxed best-match model (eq) relative to the dynamical model (dyn) for all line listed in Table~\ref{sandint3}. The vertical dashed lines in both panels indicate limits of different spectral domains.}
\label{sandinf6}
\end{figure}

In Fig.~\ref{sandinf6} we illustrate differences between line ratios of our best-match dynamic model and its corresponding equilibrium model, plotted as a function of wavelength; we present the same data in Table~\ref{sandint3}. Fig.~\ref{sandinf6}a shows that the observed line ratios are satisfactorily matched by the model (Sect.~\ref{sec:discabundances}). Equilibrium-to-dynamic model line ratios are shown in Fig.~\ref{sandinf6}b for all lines that we used with our best-match model (also including IR lines; at the assumed temperature $\teff\!\simeq\!138\,000\,$K, compare with Fig.~\ref{sandinf2} for the models $Z\ugd/10$ and $Z\ugd/100$). In this case differences can be larger than 100\% in the EUV, about 50\% in the UV, about 20--30\% in the optical wavelength range, and about 10--20\% in the infrared wavelength range. A hint of the importance of accurate line ratios is given in Fig.~\ref{sandinf3}b. If a model line ratio changes by a factor two it is easily seen that different abundances are required to match the change. Although the non-linear response to the nebula of the full model is complex, which is why plots such as Fig.~\ref{sandinf3}b are unsuitable when making quantitative estimates of abundances.

Differences in line ratios between dynamical and equilibrium models occur as a consequence of a different sensitivity of collisionally excited lines to the electron temperature. The mean temperature in the nebular region of the two models are $\langle\Te\rangle_\text{dyn}\!=\!21\,100\,$K and $\langle\Te\rangle_\text{eq}\!=\!25\,100\,$K, compare the two radial structures in Fig.~\ref{sandinf5}b, the difference is significant. The electron temperature of our evolved metal-poor models is determined by line cooling and expansion cooling. It is worth noting that although the oxygen abundance lies closer to $Z\ugd/100$ the mean model abundance is closer to $Z\ugd/10$, and it is this higher abundance that determines the physical structure of the object (see Fig.~\ref{sandinf2} and {\rP}, Figs.~15 and 16). In an observational study using the full wavelength range (EUV--infrared), where measurement errors are sufficiently small, there should be significant problems determining abundances of metal-poor objects using models that are unable to account for dynamical effects.

\section{Conclusions}\label{sec:conclusions}
{\obj} is an extraordinary object as it is a metal-poor PN with the lowest oxygen abundance known. In order to clarify contradictory abundance determinations of {\obj} in the literature we made a new study of this object using a two-fold approach. At first we re-observed the nebula and could measure a more accurate spectrum in the visual wavelength range than has been done so far. Unlike previous observational studies we could only measure an upper limit of $[\ion{Ne}{iii}]\,\lambda3869$ of $0.01\text{H}\beta$, although we measured five new lines in the nebula. We therefore chose to base our estimate of the stellar effective temperature using supplementary UV-data of Jacoby (priv.\ comm.).

In the second part of our study we used a newly calculated set of our radiation hydrodynamic models in order to determine abundances and study the influence of time-dependent effects. In this case such effects are found to be important, causing lower electron temperatures in the nebula. Resulting line strengths of dynamical models are lower than in the corresponding equilibrium models (these models are relaxed after all time-dependent terms are set to zero). Consequently, different abundances are required to match line strengths when using either approach. We found that it is only possible to make a self-consistent abundance determination using a dynamical model that is constrained using measurements in the entire wavelength range. Our final set of abundances of the five most abundant elements is: 1/1.45 (He), 1/9.8 (C), 1/8.3 (N), 1/81 (O), and 1/11 (Ne), all with respect to the mean abundance of objects in the Galactic disk ($Z\ugd$). The total metallicity is $Z\ugd/13$. Additionally, using our single $0.595\,M_\odot$ evolutionary track we found an effective temperature of {\obj} of $\teff\!\simeq\!138\,000\,$K, and a distance of $d\!=\!18\,$kpc. This distance is about the double value assumed by {\rPT}, but agrees well with the value of {\rTo}. The mean electron temperature of our dynamical model is $\langle\Te\rangle\!=\!21\,000\,$K; this is 4000\,K lower than the value of the corresponding equilibrium model.

Although we believe that our approach provides a most significant improvement when determining abundances of metal-poor objects our modeling can be improved to provide more accurate values. At first one could consider to iterate more dimensions of the parameter space, such as e.g.\ the mass of the central star and properties of the AGB wind. Three additional suggestions for such improvements that are all considered by {\rPT} are: using non-CaseB radiative transfer, replacing the black-body model of the central star with a model atmosphere, and using improved collision rates. Observationally an accurate multi-wavelength study including the infrared wavelength range, such as is announced by {\rS}, will help to constrain the models further. Last, but not least, it is important to clarify the parameters and evolutionary history of the ionizing central star(s) unambiguously.

\begin{acknowledgements}
C.\ S.\ acknowledges support by DFG grant SCHO 394/26. We thank G.\ Jacoby both for providing us with UV data prior to their publication, and for providing us feedback on a late version of the manuscript.
\end{acknowledgements}

\bibliographystyle{aa}

\Online
\begin{appendix}

\section{Intensity evolution of our RHD models}\label{sec:appendix}
\begin{figure*}[t]
\centering
\includegraphics[width=.7\textwidth,height=24cm,angle=180]{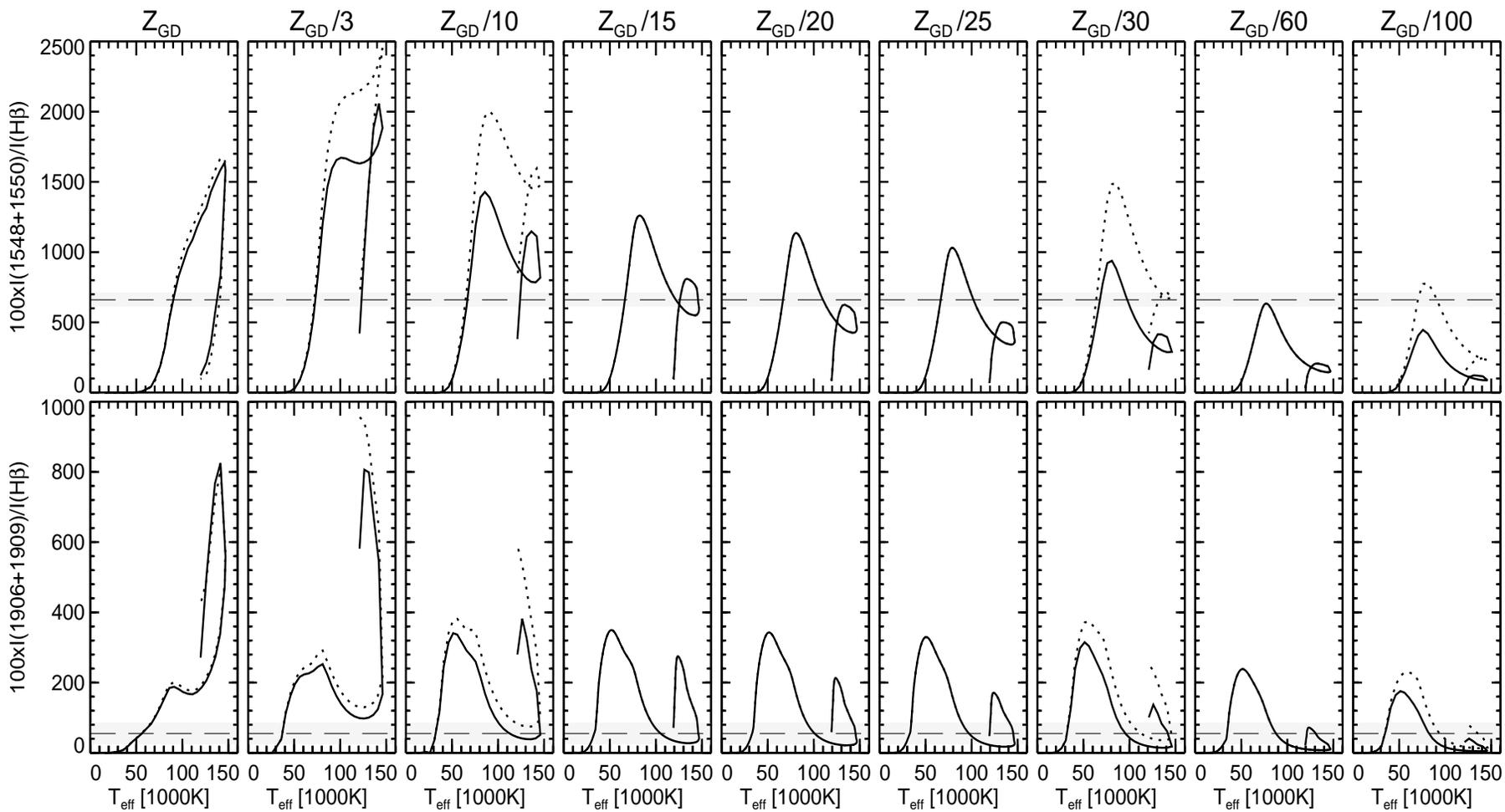}
\caption{The line strength evolution of different carbon lines as a function of {\teff} for all models.}
\label{sandinf7}
\end{figure*}

\begin{figure*}[t]
\centering
\includegraphics[width=.7\textwidth,height=24cm,angle=180]{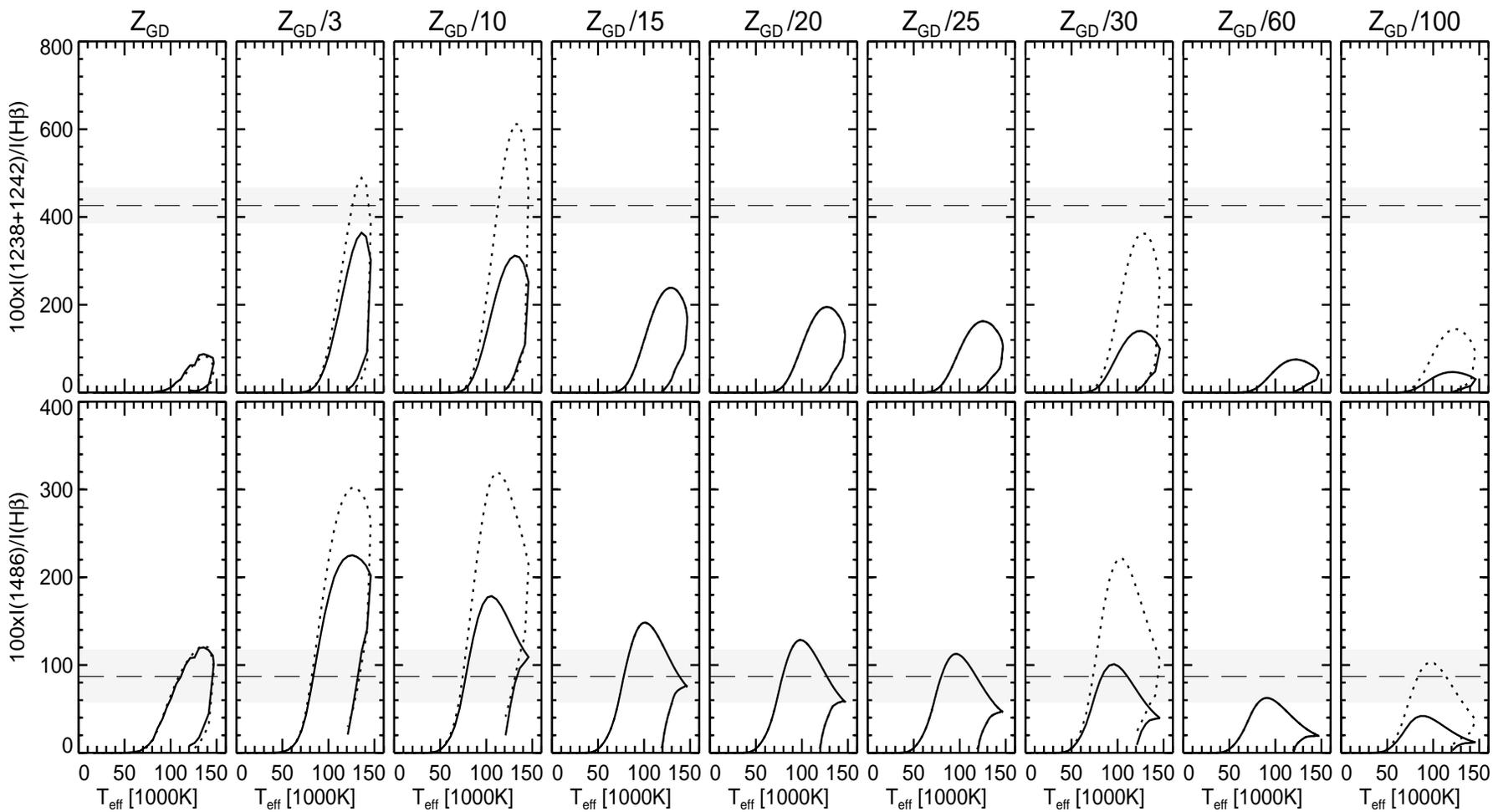}
\caption{The line strength evolution of different nitrogen lines as a function of {\teff} for all models.}
\label{sandinf8}
\end{figure*}

\begin{figure*}[t]
\centering
\includegraphics[width=\textwidth,height=24cm,angle=180]{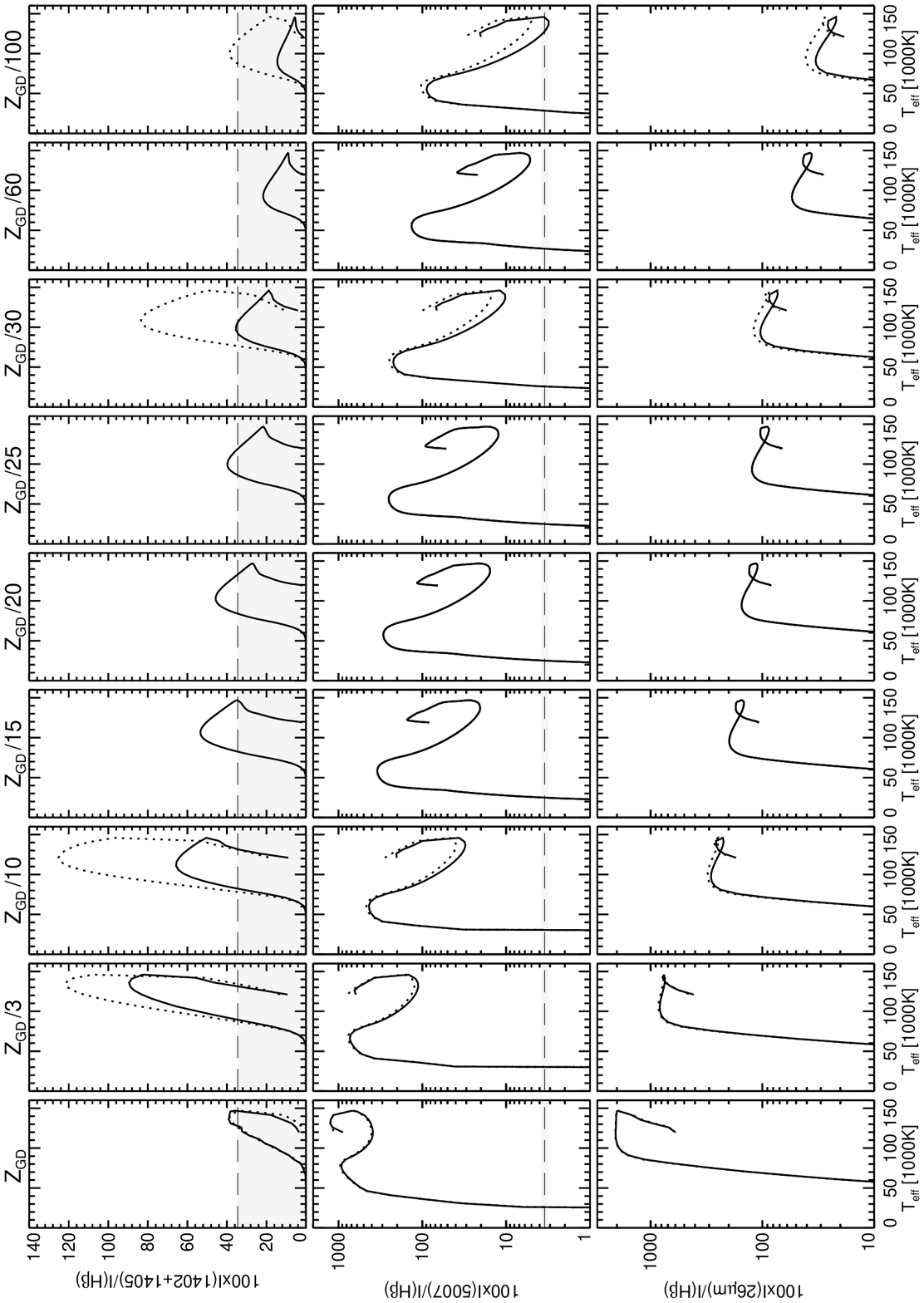}
\caption{The line strength evolution of different oxygen lines as a function of {\teff} for all models.}
\label{sandinf9}
\end{figure*}

\begin{figure*}[t]
\centering
\includegraphics[width=\textwidth,height=24cm,angle=180]{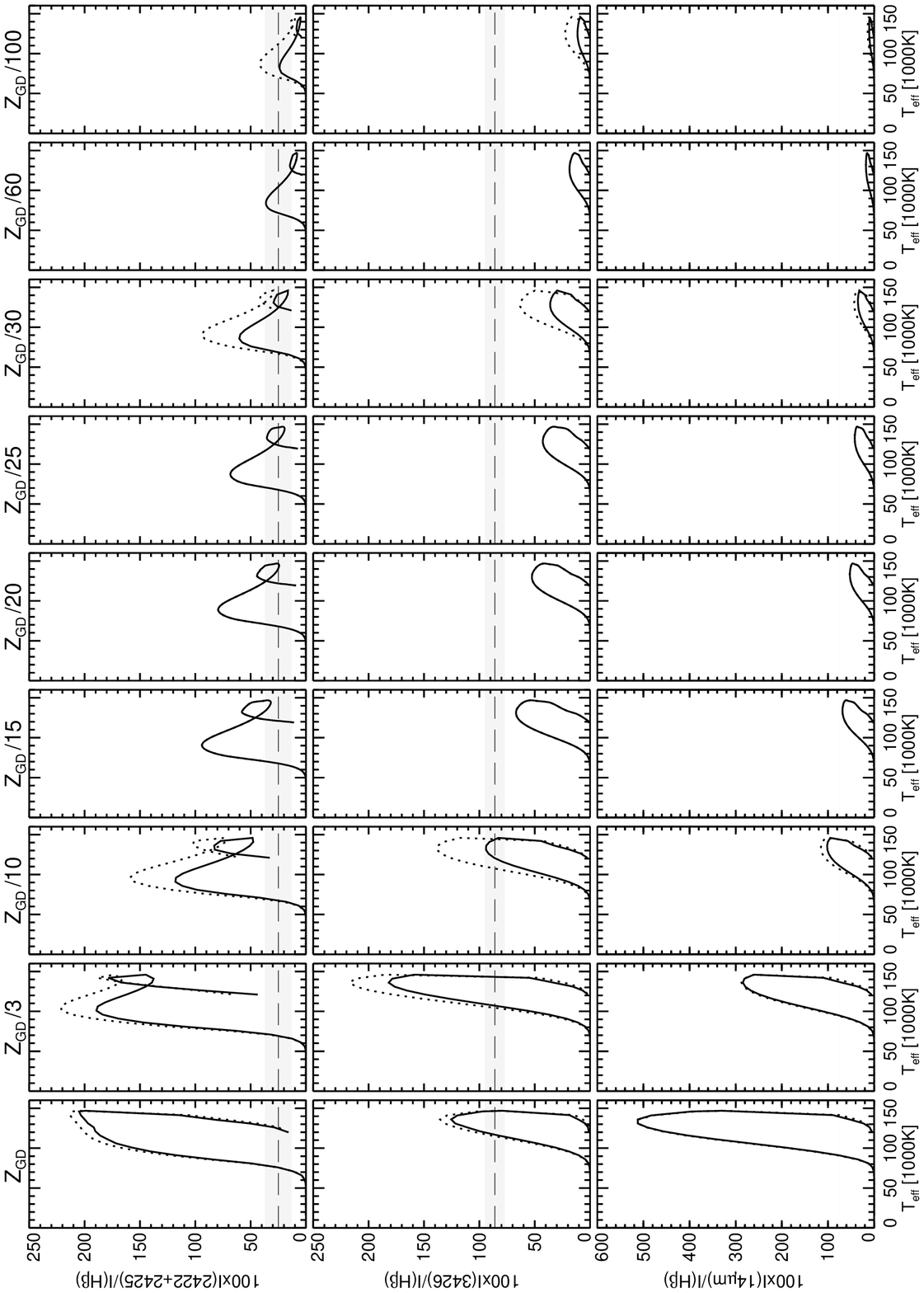}
\caption{The line strength evolution of different neon lines as a function of {\teff} for all models.}
\label{sandinf10}
\end{figure*}

\begin{figure*}[t]
\centering
\includegraphics[width=.4\textwidth,height=24cm,angle=180]{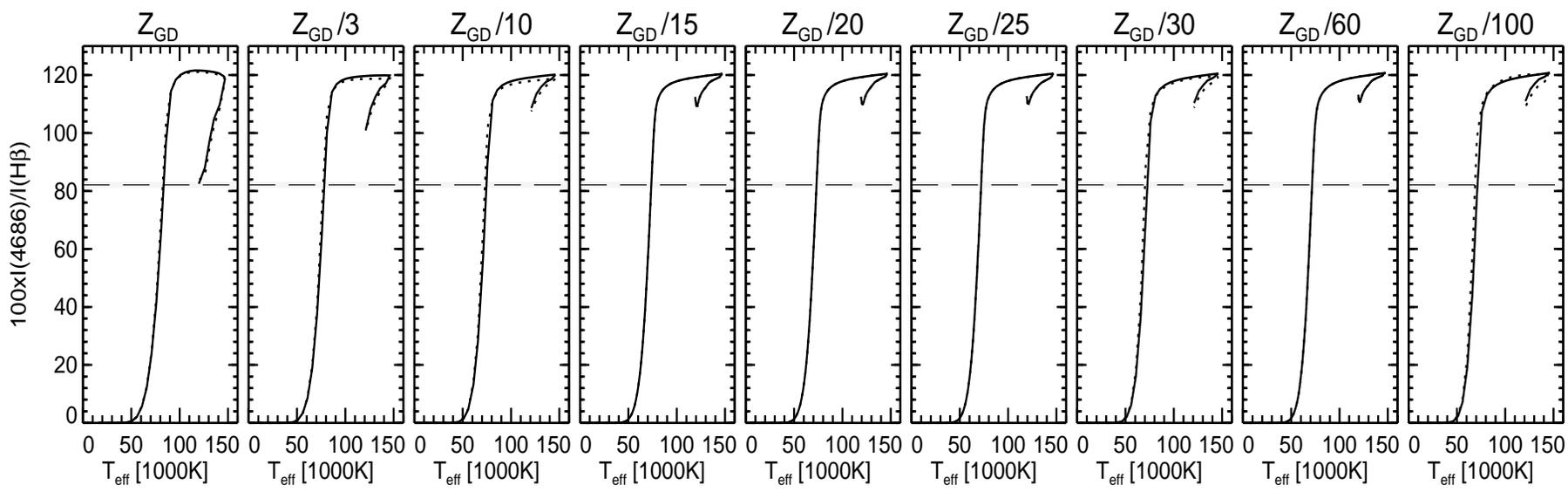}
\caption{The line strength evolution of $\ion{He}{ii}\,\lambda\,4686$ as a function of {\teff} for all models.}
\label{sandinf11}
\end{figure*}

For each model sequence we mention in Sect.~\ref{sec:discmodels} we show the intensity evolution of all emission lines of Table~\ref{sandint3} in Figs.~\ref{sandinf7}-\ref{sandinf11}. Solid lines show dynamic models and dotted lines equilibrium models (for those sequences where they were calculated). The abscissa is in every case the stellar effective temperature \teff. Horizontal dashed lines mark observed values for {\obj}, and gray shaded regions mark corresponding error intervals. Additionally we show the evolution of the two infrared lines, $[\ion{O}{iv}]\,\lambda26\,\mu$m and $[\ion{Ne}{v}]\,\lambda14\,\mu$m, despite a lack of currently existing observational data.

\end{appendix}
\end{document}